\documentclass[prd,a4paper,preprint,preprintnumbers,nofootinbib,superscriptaddress]{revtex4-2}

\usepackage[a4paper, hdivide={1.91cm,,1.165cm}, vdivide={1.83cm,,3.0cm}]{geometry}

\usepackage{amsmath}

\usepackage{colortbl}
\usepackage{bbold}

\usepackage{amsfonts}

\usepackage{graphicx}
\usepackage{xspace}
\usepackage{color}
\usepackage{units}
\usepackage{slashed}
\usepackage[hyperfootnotes=false]{hyperref}
\hypersetup{linkcolor=red,
citecolor=green,
filecolor=cyan,
urlcolor=magenta}

\usepackage{gensymb}
\usepackage{tabularx}
\usepackage{booktabs}
\usepackage{multirow}
\usepackage{xcolor}
\usepackage{setspace}
\usepackage{orcidlink}
\usepackage{physics}
\usepackage{adjustbox}

\usepackage{enumitem}

\newcommand{\beq}{\begin{eqnarray}}
\newcommand{\eeq}{\end{eqnarray}}

\def\ds{\displaystyle}
\def\beq{\begin{equation}}
\def\eeq{\end{equation}}
\def\bea{\begin{eqnarray}}
\def\eea{\end{eqnarray}}
\def\beeq{\begin{eqnarray}}
\def\eeeq{\end{eqnarray}}
\def\ve{\vert}
\def\vel{\left|}
\def\ver{\right|}
\def\nnb{\nonumber}

\def\lla{\left<}
\def\rra{\right>}

\def\nnb{\nonumber}

\def\ba{\begin{array}}
\def\ea{\end{array}}

\def\xis0{{\Xi^{*0}}}

\def\g5{\gamma_5}

\def\es{&=& }

\def\ar{&+& }
\def\ek{&-&}

\def\cp{&\times& }

\allowdisplaybreaks[1]

\begin{document}


\title{The study of weak decays induced by $\frac{1}{2}^+ \to \frac{3}{2}^-$ transition in light-cone sum rules}

\author{T.~M.~Aliev\,\orcidlink{0000-0001-8400-7370}}
\email{taliev@metu.edu.tr}
\affiliation{Department of Physics, Middle East Technical University, Ankara, 06800, Turkey}

\author{S.~Bilmis\,\orcidlink{0000-0002-0830-8873}}
\email{sbilmis@metu.edu.tr}
\affiliation{Department of Physics, Middle East Technical University, Ankara, 06800, Turkey}
\affiliation{TUBITAK ULAKBIM, Ankara, 06510, Turkey}

\author{M.~Savci\,\orcidlink{0000-0002-6221-4595}}
\email{savci@metu.edu.tr}
\affiliation{Department of Physics, Middle East Technical University, Ankara, 06800, Turkey}


\begin{abstract}
In this study, we analyzed the weak decays induced by $J^P = \frac{1}{2}^{+} \to  \frac{3}{2}^{-} $ transitions within the light-cone sum rules. Specifically, semileptonic decays of the bottom baryons into the P-wave baryons $\Lambda_b \to \Lambda_c(2625) \ell \nu_l$ and $\Xi_b \to \Xi_c(2815) \ell \nu_l$ as well as nonleptonic $\Lambda_b \to \Lambda_c(2625) \pi (\rho)$ and $\Xi_b \to \Xi_c(2815) \pi (\rho)$ decays are investigated. The form factors for the considered transitions are obtained within the sum rules method. With the calculated form factors, the decay widths of the processes are determined. Up to now, only the decay width  for $\Lambda_b^0 \to \Lambda_c^+ \mu^- \nu_\mu$ has been measured among the considered decays, and we observe that our finding is quite compatible with the measurement. We also compare our results with the predictions of other approaches.  
\end{abstract}

\maketitle

\newpage

\section{Introduction}
The study of weak decays of bottom baryons is a promising  area in heavy flavor physics. Bottom baryons, due to their higher mass, exhibit numerous decay modes, making them an excellent testing ground for Quantum Chromodynamics (QCD) and exploring the possibility of new physics beyond the Standard Model (SM). The analysis of $b \to c$ transitions holds phenomenological importance for the precise determination of the Cabibbo-Kobayashi-Maskawa (CKM) matrix element $V_{cb}$, as well as for testing the lepton universality~{\cite{LHCb:2022qnv,LHCb:2022zom,Belle-II:2023qyd,Guadagnoli:2022oxk,Wang:2017jow,Gao:2021sav,Cui:2023bzr,Guadagnoli:2022oxk} and searching for new physics beyond the Standard Model. For these reasons, the semileptonic and nonleptonic decays of the bottom baryons to ground state
charmed baryons have been comprehensively discussed in the literature within various theoretical frameworks such as lattice QCD \cite{Gottlieb:2003yb,Detmold:2015aaa}, QCD sum rules
\cite{Huang:2005mea,Zhao:2020mod,Azizi:2018axf}, light cone QCD sum rules
\cite{Duan:2022uzm,Miao:2022bga,Wang:2009yma}, and various phenomenological models
\cite{Pervin:2005ve,Ebert:2006rp,Ke:2007tg,Faustov:2016pal,Gutsche:2015mxa,Rahmani:2020kjd,Chua:2018lfa,Ke:2019smy,Chua:2019yqh,Gutsche:2018nks,Geng:2020ofy,Li:2021qod,Guo:2005qa}.

However, the semileptonic decays of the bottom baryons into the P-wave baryons have received less attention. Nonetheless, few studies are exploring the decays of the heavy bottom baryons to the excited P-wave baryons using different approaches like the constituent quark model \cite{Pervin:2005ve}, covariant confined model 
\cite{Gutsche:2018nks},
lattice QCD \cite{Meinel:2021rbm}, and light front-quark model
\cite{Li:2021qod,Li:2022hcn}.

In this study, we focus on weak decays involving two heavy baryons, specifically analyzing two semileptonic decays, $\Lambda_b \to \Lambda_c(2625) \ell \nu$ and $\Xi_b \to \Xi_c(2815) \ell \nu$, as well as nonleptonic decays, $\Lambda_b \to \Lambda_c(2625) M$ and $\Xi_b \to \Xi_c(2815) M$, where $M$ represents the $\pi$ or $\rho$ meson. It is important to note that experimental data is only available for the $\Lambda_b \to \Lambda_c(2625) \ell \nu$ decay. The objectives of this work are twofold. First, we aim to compare the theoretical results for the $\Lambda_b \to \Lambda_c(2625) \ell \nu$ decay with the available experimental data. Second, we aim to provide theoretical predictions for the branching ratios of the semileptonic $\Xi_b \to \Xi_c(2815) \ell \nu$ decay and the nonleptonic $\Lambda_b \to \Lambda_c(2625) M$ and $\Xi_b \to \Xi_c(2815) M$ decays, which have not yet been measured but have the potential of being discovered with the upgraded LHCb. One of the main challenges in the theoretical studies of these decays is as follows: The interpolating current for $J = \frac{3}{2}$ baryon not only interacts with positive and negative parity $J = \frac{3}{2}$ baryons but also with $J^P = \frac{1}{2}^-$ baryon. This interaction leads to the unwanted contributions. To remove this pollution, the combinations of the sum rules corresponding to different Lorentz structures are constructed.

The work is organized as follows: In Sec~\ref{sec:2}, the sum rules for the relevant form factors are derived. In Sec.~\ref{sec:3}, the numerical analysis is performed to determine the form factors, and with the obtained results, the decay widths of the considered transitions are estimated. The last section contains our conclusion.
\section{The light-cone sum rules of $H_b \to H_c $ transition form factors}
\label{sec:2}
For the sake of simplicity, we will denote $\Xi_b(\Lambda_b)$ and $\Xi_c(\Lambda_c)$ baryons as $H_b$ and $H_c$, respectively, for further discussions. The transition form factors for $H_b(\frac{1}{2}^+) \to H_c(\frac{3}{2}^+)$ induced by the $V-A$ current $\bar{c} \gamma_\mu (1-\gamma_5)b$ are defined as
\cite{Chua:2019yqh},
\bea
   \label{eqn:1}
     \lla H_c (p^\prime,s) \ve \bar{c}\gamma_\mu(1 - \gamma_5)b 
\ve H_b(p,s)\rra \es
\bar{u}_\alpha(p^\prime,s)\Bigg\{ \Bigg[\frac{p_\alpha}{m_1}\Bigg(\gamma_\mu F_1^+ 
+ \frac{p_\mu}{m_1}F_2^+ 
+ \frac{p^\prime_\mu}{m_+}F_3^+ \Bigg) + g_{\mu\alpha} F_4^+ \Bigg]\gamma_5 \nnb \\
\ek \Bigg[ \frac{p_\alpha}{m_1} \Bigg (\gamma_\mu G_1^+ +
\frac{p_\mu}{m_1}G_2^+ + \frac{p^\prime_\mu}{m_+} G_3^+ \Bigg) +
g_{\mu\alpha} G_4^+ \Bigg] \Bigg\} u(p,s)\,,
\eea
where $u_\alpha (p^\prime)$ and $u(p)$ are the Rarita-Schwinger and Dirac
spinors, $m_1$ and $m_+$ are the masses of the initial
and final baryons, respectively, and $F_i^+$ and $G_i^+$ are the form factors.
The matrix element for $H_b(\frac{1}{2}^+) \to H_c(\frac{3}{2}^-)$ transition
can be obtained from Eq. (\ref{eqn:1}) by making the following replacements: 
$F_i^+ \to F_i^-$, $G_i^+ \to G_i^-$, $\gamma_5 \to 1$, $1 \to
\gamma_5$, and $m_+ \to m_-$, where $m_-$ 
is the mass of the $H_c(\frac{3}{2}^-)$ baryon.

To calculate the form factors responsible for the $H_b(\frac{1}{2}^+) \to H_c(\frac{3}{2}^-)$ transitions, we start with the following vacuum to $H_b$
baryon correlation function.
\bea
   \label{eqn:2}
   \Pi_{\mu\nu} (p,q) = i \int d^4x e^{ip^\prime x} \lla 0 \ve 
T \{\eta_{\mu}(x) j_\nu (0)\} \ve H_b(p)\rra\,,
\eea
Here $\eta_{\mu}$ is the interpolating current of the final heavy
$H_c$ baryon and $j_\nu (0) = \bar{c} \gamma_\nu (1-\gamma_5)b$
is the current describing the $b \to c$ transition.

The interpolating current for the heavy $H_c$ baryons is chosen in the following form 
\cite{Wang:2017vtv} 
 \bea
    \label{eqn:3}
   \eta_\mu = \epsilon^{abc} \Big[\partial_\alpha \partial_\beta
\Big(q_1^{aT} C \gamma_5 q_2^b \Big) \Big]\Gamma_{\alpha \beta \mu}
c^c\,,
\eea
where
\bea
\Gamma_{\alpha \beta \mu} = \Bigg( g_{\mu \alpha} g_{\beta \rho} +
g_{\alpha \rho} g_{\beta \mu} - \frac{1}{2} g_{\alpha \beta}
g_{\mu \rho}\Bigg) \gamma^\rho \gamma_5\,,\nnb
\eea
$a$, $b$, and $c$ are the color indices, and $C$ is the charge conjugation operator.

The reason for choosing the current in this form is as follows:
In this form, the relative angular momentum of the diquark is $L_\rho = 0$ , i.e., it is in S-wave. On the other hand, the angular momentum between the light diquark and heavy quark is equal to $L_{\lambda} = 2$, which is achieved by applying two consecutive derivatives.

It should be noted that the interpolating current $\eta_{\mu} (x)$
couples not only with $J^P = \frac{3}{2}^+$ state but also to the
heavy baryon with negative parity $J^P = \frac{3}{2}^-$. Hence, the hadronic part of the correlation
function is modified by  the contribution of the negative parity resonance.

We proceed with our analysis by calculating the correlation function from the
phenomenological (hadronic) side. At the hadronic level, the correlation 
function is obtained by sandwiching the currents between hadronic states,
\bea
\label{eqn:4}
\Pi_{\mu\nu} \es \frac{1}{ m_{+}^2 - p^{\prime 2} } \lla 0 \ve
\eta_\mu \ve H_c^+ \rra \lla H_c^+ \ve \bar{c}\gamma_\mu
(1-\gamma_5) b \ve H_b \rra\,, \nnb \\
\ar \frac{1}{ m_{-}^2 - p^{\prime 2} } \lla 0 \ve \eta_\mu \ve
H_c^- \rra \lla H_c^- \ve \bar{c} \gamma_\mu (1-\gamma_5) b \ve
H_b (p) \rra + \cdots 
\eea
where dots represent the contributions of excited states and continuum,
$H_c^{+(-)}$ is positive (negative) parity heavy baryon. The matrix elements
$\lla 0 \ve \eta_\mu \ve H_c^+ \rra$ and $\lla 0 \ve \eta_\mu \ve
H_c^- \rra$ are defined as
\bea
\label{eqn:5}
\lla 0 \ve \eta_\mu \ve H_c^+ \rra \es \lambda_{+} u_\mu(p^\prime)~, \nnb \\
\lla 0 \ve \eta_\mu \ve H_c^-  \rra \es \lambda_{-} \gamma_5 u_\mu(p^\prime)~.
\eea

Using Eqs. (\ref{eqn:4}) and (\ref{eqn:5}) and performing summation over spins
of Rarita-Schwinger spinors using
\bea
\label{eqn:6}
 \sum u_\mu (p^\prime,s) \bar{u}_\beta (p^\prime) = -(p^\prime+m) 
\Bigg[g_{\mu \beta}-\frac{1}{3}
\gamma_\mu \gamma_\beta - \frac{2}{3} \frac{p^\prime_\mu p^\prime_\beta}{m^2}+\frac{1}{3}
\frac{p^\prime_\mu \gamma_\beta - p^\prime_\beta \gamma_\mu}{m} \Bigg]\,,
\eea
one can calculate the hadronic part of the correlation function.

We would like to make the following preliminary remarks before proceeding with further calculations. 

\begin{itemize}[noitemsep]
    \item[a)] The interpolating current of the spin $\frac{3}{2}$
baryon has nonzero overlap not only with 
$J=\frac{3}{2}$ state, but also with $J^P = \frac{1}{2}$ baryon. Indeed,
\bea
\lla 0 \vel \eta_\mu \ver \frac{1}{2} (p^\prime) \rra
\sim (\gamma_\mu - \frac{4 p^\prime_\mu}{m}) u(p^\prime,s)\,. \nnb
\eea
    It follows from this expression that the Lorentz structures
$\gamma_\mu$ or $p^\prime_\mu$ contain contributions of spin-$\frac{1}{2}$ baryons. Based on Eq. (\ref{eqn:6}), we can infer that the structure $g_{\mu \beta}$ solely contains spin-3/2 baryon contributions, excluding any spin-1/2 baryon contributions.

\item[b)] From Eqs. (\ref{eqn:5}) and (\ref{eqn:6}), we observe
that the hadronic part of the correlation function contains
many Lorentz structures. However, not all these structures are independent.

To obtain the independent structures, a specific order of
Dirac matrices are needed to be specified. In the present work, we choose
$\gamma_\mu \rlap/q \gamma_\nu ( \gamma_5)$.
\end{itemize}

Taking these remarks into account, we get the following results for the correlation function from the hadronic side.
\bea
\label{eqn:7}
\Pi_{\mu \nu} \es \frac{\lambda_+}{m_+^2 - p'^2}(\not\!p^\prime + m_+)
\Bigg\{\frac{p_\mu}{m_1}\Bigg(\gamma_\nu F_1^+ + \frac{p_\nu}{m_1}F_2^+
+ \frac{p'_\mu}{m_+}F_3^+ \Bigg )+ g_{\mu \nu} F_4^+ \Bigg] \gamma_5 \nnb \\
\ek  \Bigg[\frac{p_\mu}{m_1}\bigg(\gamma_\nu G_1^+ + \frac{p_\nu}{m}
G_2^+ + \frac{p'_\nu}{m_+} G_3^+ \Bigg )+g_{\mu \nu} G_4^+ \Bigg] \Bigg\}u(p) \nnb \\
~ \ar \frac{\lambda_{-} \gamma_5}{m_{-}^2 - p'^2}(\not\!p^\prime +
m_{-}) \Bigg\{ \Bigg [\frac{p_\mu}{m_1} \Bigg(\gamma_\nu F_1^- + \frac{p_\nu}{m_1}F_2^-
+ \frac{p'_\nu}{m_{-}}F_3^- \Bigg)+g_{\mu \nu} F_4^- \Bigg] \nnb \\
\ek  \Bigg[\frac{p_\mu}{m_1} \Bigg(\gamma_\nu G_1^- + \frac{p_\nu}{m_1} G_2^- +
\frac{p'_\nu}{m_-}G_3^- \Bigg) + g_{\mu \nu}G_4^- \Bigg]\gamma_5  \Bigg\} u(p) \,.
\eea
Applying the Dirac equation and replacing the four-momentum of $H_b$ baryon
with $p_\mu \to m v_\mu$ in Eq. (\ref{eqn:7}), where $v_\mu$ is its velocity, the above equation takes the following form,
\bea
\label{eqn:8}
  \Pi_{\mu \nu} \es \frac{\lambda_+}{m_{+}^2 - p^{\prime 2}}
\Bigg\{v_\mu \Big[ F_1^+ \Big(2m_1 v_\nu + (m_1 + m_+) \gamma_\nu
- \rlap/q \gamma_\nu \Big) \gamma_5 + F_2^+ v_\nu \Big[
    (m_+ - m_1) - \rlap/q\Big] \gamma_5 \nnb \\
\ar F_3^+ \frac{(m_1 v_\nu - q_\nu)}{m_+} \Big[ - \rlap/q
+ (m_+ - m_1) \Big] \gamma_5 \bigg] +
      F_4 g_{\mu \nu} ( -m_1 + m_+ - \rlap/q) \gamma_5 \nnb \\
\ek v_\mu \Big[ G_1^+ \Big( 2 m_1 v_\nu + (m_+ - m_1) \gamma_\nu
- \rlap/q \gamma_\nu \Big) + G_2^+ v_\nu \bigg( (m_1 + m_+) - \rlap/q \Big) \nnb  \\
\ar
  G_3^+ \frac{m_1 v_\nu - q_\nu}{m_+} \Big( (m_+ + m_1)
- \rlap/q \Big) \Big] -
      G_4^+ g_{\mu \nu} \big(m_+ + m_1 - \rlap/q \big) \bigg\} u(p) \nnb \\
\ar \Big( \lambda_+ \to \lambda_{-}, m_+ \to - m_{-},
F_1^+ \to F_1^-, F_2^+ \to -F_2^-, F_3^+ \to -F_3, F_4^+ \to - F_4^-, \nnb \\
&& G_1^+ \to G_1^-, G_2^+ \to G_2^-, G_3^+ \to G_3^-,
G_4^+ \to G_4^- \Big)\,.
\eea
In the present study, we analyze the semileptonic decays of positive parity
spin-$1/2$ $H_b$ baryon to spin-$3/2$ baryon with negative parity
$H_c$. Since this transition is described
by the form factors $F_i^-(q^2)$ and $G_i^-(q^2)$,  the contribution of the form factors $F_i^+(q^2)$ and $G_i^+(q^2)$
should be eliminated. For this goal, the combinations of the sum rules obtained from the different Lorentz structures should be considered. The method of eliminating the unwanted pollution by considering the linear combinations of sum rules obtained from different Lorentz structures was proposed in~\cite{Khodjamirian:2011jp}.

Having obtained the correlation function from the hadronic side, let us turn our attention to the calculation of it given in Eq. (\ref{eqn:2}) in terms of quarks and gluons using the operator product expansion. After applying  Wick theorem for the
correlation function, we get
\bea
\label{eqn:9}
\Pi_{\mu \nu} \es i \int d^4 x e^{i p^\prime x}\epsilon^{abc}
(C\gamma_5)_{\phi \eta} (\Gamma_{\alpha\beta\mu})_{\rho\gamma}
[\gamma_\nu (1-\gamma_5)]_{h \xi}
S_{\gamma h} (x) \nnb \\
\cp \partial_\alpha\partial_\beta \Big(\lla 0 \ve
q_{1\phi}^a (x) q_{2\eta}^b (x)
b_\xi^c (0) \ve H_b (p) \rra\Big)\,,
\eea
where $S(x)$ is the c-quark propagator. The matrix element, $\epsilon^{abc}\lla 0\ve q_{1\phi}^a (x)
q_{2\eta}^b (0) b_\xi^c (0) \ve H_b (p) \rra$ must be determined to calculate the theoretical part of the correlation function from the QCD side. This matrix element is expressed in terms of the $H_b$ baryon distribution amplitudes
(DAs) whose explicit forms  are given in \cite{Ali:2012pn,Bell:2013tfa,Ball:2008fw}. When the polarization vector is parallel to the light cone plane, the matrix element mentioned above is parametrized in terms
of the four DAs,%
\bea
    \label{eqn:10}
    \epsilon^{abc} \langle 0\ve s_\alpha ^a (t_1) q_\beta ^b (t_2)
b_\sigma ^c (0) \ve H_b (v) \rangle = \sum_{i=1}^4 \Lambda_i
(\Gamma_i)_{\beta \alpha} u(v)_\sigma
\eea
where
\bea
\begin{aligned}
    \label{eqn:11}
        \Lambda_1 &= \frac{1}{8} v_+ f^{(1)} \psi_2 \,,
& \Gamma_1 &= \gamma_5 C^{-1}\,, \\
        \Lambda_2 &= \frac{1}{8}f^{(2)}\psi_3^\sigma \,,
 & \Gamma_2 &= i \sigma_{\rho \beta} n_\rho \bar{n}_\beta \gamma_5 C^{-1}\,, \\
        \Lambda_3 &= \frac{1}{4} f^{(2)} \psi_3^{(s)} \,,
& \Gamma_3 &= \gamma_5 C^{-1}~ \\
        \Lambda_4 &= - \frac{1}{8 v_+} f^{(1)} \psi_4 \,,
& \Gamma_4 &= \gamma_5 C^{-1}~
\end{aligned}
\eea
and
 \bea
     n_\mu = \frac{x_\mu}{vx} \,,~~ \Bar{n}_\mu = 2 v_\mu - n_\mu  \,, \nnb
 \eea
$f^{(1)}$ and $f^{(2)}$ are the decay constants and $\psi_i$ are the DAs of $H_b$ baryon with definite twist.

The light cone distribution amplitudes are scale-dependent functions. To obtain their scale dependency, it is convenient to go to momentum representation with the help of the Fourier transformation
\begin{equation}
  \label{eq:2}
  \begin{split}
    \psi(t_1,t_2) &= \int_0^{\infty} d\omega_1 \int d\omega_2 e^{-i \omega_1 t_1 -i \omega_2 t_2 } \psi(\omega_1, \omega_2) \\
                 &= \int_0^\infty \omega d\omega \int_0^1 du e^{-i \omega (t_1 u + t_2 \bar{u})} \psi(\omega,u)
  \end{split}
\end{equation}
In the first expression, $\omega_1$ and $\omega_2$ are the energies of the light quarks. In the second expression, instead of $\omega_1$ and $\omega_2$, new variables $\omega$ and $u$, where $\omega = \omega_1 + \omega_2$ is the total energy carried by light quarks, and dimensionless variable $u$ corresponds to the relative energy carried by light quark $q_1$, have been introduced. Moreover, using the definition $t_1 = v x_i$, where $t_i$ is the distance between $i^{th}$ light quark and the origin along the light-cone vector $n$ direction, for the distribution amplitudes, we get,
\begin{equation}
  \label{eq:3}
  \psi(t_1,t_2) = \int_0^\infty d \omega \omega \int_0^1 du e^{-i\omega v [u x_1 + \bar{u} x_2]} \psi(\omega,u),
\end{equation}
when $t_1 = t_2$, we obtain,
\begin{equation}
  \label{eq:4}
  \psi(t,t) = \int_0^\infty d\omega \omega \int_0^1 du e^{-i \omega v x} \psi(\omega,u)
\end{equation}
This equation will be used in further calculations.

The DAs of $H_b$ baryon, which we will use in further
analysis, are obtained in \cite{Ali:2012pn} as follows: 
\bea
    \label{eqn:13}
      \psi_2 (\omega, u) \es \omega^2 u \bar{u} \sum_{n=0}^{2}
\frac{a_n}{\epsilon_n^4} \frac{C_n^{3/2}(2 u -1)}{|C_n^{3/2}|^2}
e^{-\omega/\epsilon_n}\,, \nnb \\
      \psi_3 (\omega, u) \es \frac{\omega}{2}  \sum_{n=0}^{2}
\frac{a_n}{\epsilon_n^3} \frac{C_n^{1/2}(2 u -1)}{|C_n^{1/2}|^2}
e^{-\omega/\epsilon_n}\,, \nnb \\
      \psi_4(\omega, u) \es  \sum_{n=0}^{2} \frac{a_n}{\epsilon_n^2}
\frac{C_n^{1/2}(2 u -1)}{|C_n^{1/2}|^2} e^{-\omega/\epsilon_n}\,,
\eea
where the sub-index of $\psi$ indicates the twist of the distribution
amplitudes, $C_n^\lambda$ is the Gegenbauer polynomials, and
$|C_n^\lambda| = \int_0^1 du |C_n^\lambda (2u -1)|^2$.
The parameters appearing in DAs of $\Xi_b$ and $\Lambda_b$  are presented
in Table \ref{tab:1} for completeness.

\begin{table}[hbt]
\begin{adjustbox}{width=\columnwidth,center}
  \renewcommand{\arraystretch}{1.4}
\setlength{\tabcolsep}{8pt}
\begin{tabular}{c|c|ccc|ccc}
\toprule
\multirow{5}{*}{$\Lambda_b$}  & twist & $a_0$ & $a_1$ & $a_2$ & $\varepsilon_0(\mathrm{GeV})$                         &
$\varepsilon_1(\mathrm{GeV})$ & $\varepsilon_2(\mathrm{GeV})$ \\
\cline { 2 - 8 }   & 2        & 1     & $-$   & $\frac{6.4 (1-A)}{1.44-A}$      & $\frac{2 - 1.4 A}{6.7-A}$           &
$-$                & $\frac{0.32 (1-A)}{0.83 - A}$            \\
                   & $3 s$    & 1     & $-$   & $\frac{0.04 (1 - 3 A)}{-0.4 - A}$ & $\frac{0.21 + 0.56 A}{1.6 + A}$    &
$-$                & $\frac{ 0.09 - 0.25 A}{1.41 - A}$        \\
                   & $3 a$    & $-$   & 1     & $-$                             & $-$                                 &
$\frac{ 0.08 + 0.35 A}{0.2 + A}$   & $-$                      \\
                   & 4        & 1     & $-$   & $\frac{-0.12 + 0.07 A}{1.34 - A}$ & $\frac{-0.87 + 0.65 A}{-2 + A}$   &
$-$                & $\frac{-9.3 + 5.5 A}{-30 + A}$          \\
  \midrule
  \midrule
  \multirow{5}{*}{$\Xi_b$}    & twist & $a_0$ & $a_1$ & $a_2$ & $\varepsilon_0(\mathrm{GeV})$ &
$\varepsilon_1(\mathrm{GeV})$               & $\varepsilon_2(\mathrm{GeV})$               \\
\cline { 2 - 8 }   & 2        & 1     & $\frac{0.71 - 0.25 A}{1.68 - A}$     & $\frac{7.2 - 6.6 A}{1.68 - A}$      &
$\frac{2.4 - 1.4 A}{7.7 - A}$ & $\frac{-1.67 + 0.57 A}{-5 + A}$              & $\frac{0.39 - 0.36 A}{0.98 - A}$   \\
                   & $3 s$    & 1     & $\frac{0.1 + 0.04 A}{0.6 + A}$       & $\frac{0.03 (1  - 4 A)}{ -0.6 - A}$  &
$\frac{0.35 + 0.56 A}{1.9 + A}$       & $\frac{ 65 + 27 A}{160}$             & $\frac{0.06 (1 - 5 A)}{1.54 - A}$   \\
                   & $3 a$    & $\frac{0.16 (2 - A)}{-0.3 - A}$             & 1 & $\frac{0.17 A}{ 0.3 + A}$        &
$       0.11         $        & $\frac{0.1 + 0.39 A}{0.3 + A}$               & $ 0.33 $                            \\
                   & 4        & 1     & $\frac{0.14 - 0.03 A}{1.16 - A}$     & $\frac{-0.13 + 0.1 A}{1.61 - A}$     &
$\frac{1.01 - 0.63 A}{2.3 - A}$    & $\frac{3.92 - 0.82 A}{2.9 + A}$      & $\frac{ 1.54 - 1.2 A}{5.1 - A}$        \\
\bottomrule
\end{tabular}
\end{adjustbox}
  \caption{Parameters appearing in the DAs of the form factors. In our calculations we used $A=\frac{1}{2}$.}
  \label{tab:1}
\end{table}

Using Eqs.(\ref{eqn:9},\ref{eqn:10}) and (\ref{eq:4}), we get
\bea
    \Pi_{\mu \nu} \es i \int d^4 x
\int du ~ \Bigg(\partial_\alpha \partial_\beta
\int d\omega \omega e^{-i \omega v x} \Bigg) \nnb \\
\cp \Bigg\{\sum \Lambda_i {\rm Tr}(C \gamma_5 \Gamma_i) \Gamma_{\alpha \beta \mu}
S(x) \gamma_\nu (1-\gamma_5) \Bigg\}u(v)\,,\nnb
\eea
where $S(x)$ is the $c$-quark propagator given as,
\bea
    S_(x) = \int d^4 k e^{-ikx}\frac{i(\not\!k+m_c)}{k^2 - m_c^2}\,.\nnb
\eea
Performing integrations over $x$ and $k$, the expression of the
correlation function from the QCD side is obtained as follows:
\bea
  \label{eqn:14}
    \Pi_{\mu \nu} \es \int du \int d\omega ~\omega^3
{2 f^{(2)} \psi_{3s} \over  \bar{\sigma}  (p^{\prime 2} -s)}
\Bigg\{2 \rlap/{q} (1 - \gamma_5) v_{\mu} v_{\nu} -
\rlap/{q} \gamma_{\nu}  (1 + \gamma_5) v_{\mu} \nnb \\
\ar m_c \gamma_{\nu} (1 + \gamma_5) v_{\mu}
- 2 m_c (1 - \gamma_5) v_{\mu} v_{\nu} -
\gamma_{\nu} (1 - \gamma_5) v_{\mu}
  {(q^2 + \omega m_1 - s) \over m_1}\Bigg\}\,.
\eea
where,
\begin{equation}
  \label{eq:1}
  \begin{split}
    s(\sigma) = \frac{ m_c^2 + \sigma (m_1^2 - q^2) - \sigma^2 m_1^2}{\bar{\sigma}}\,,
  \end{split}
\end{equation}
in which $\bar{\sigma} = 1 - \sigma$ and $\sigma = \frac{\omega}{m_1}$.

Equating the coefficients of the structures from theoretical and hadronic parts of the correlation function and combining the sum rules obtained for different Lorentz structures, and performing the Borel transformation over the variable $-p^{\prime 2}$, we get the following sum rules for the form factors $F_i^-$ and $G_i^-$,
\begin{table}[hbt]
\begin{adjustbox}{center}
\renewcommand{\arraystretch}{1.4}
\setlength{\tabcolsep}{8pt}
  \begin{tabular}{cc}
    \toprule
$\Pi_i$  & Structures \\
\midrule
$\Pi_{2}(\Pi_{10})$  &   $\gamma_5 v_\mu v_\nu ~ ( v_\mu v_\nu )$  \\
$\Pi_{3}(\Pi_{11})$  &   $\gamma_{\nu} \gamma_5 v_\mu ~ ( \gamma_{\nu} v_\mu )$  \\
$\Pi_{4}(\Pi_{12})$  &   $\not\!{q} \gamma_{\nu} \gamma_5 v_\mu ~ ( \not\!{q} \gamma_{\nu} v_\mu )$  \\
$\Pi_{6}(\Pi_{14})$  &   $\not\!{q} \gamma_5 v_\mu v_\nu ~ ( \not\!{q} v_\mu v_\nu )$  \\
    \bottomrule
  \end{tabular}
\end{adjustbox}
\caption{Invariant functions $\Pi_i$ and their corresponding structures.}
\label{tab:2}
\end{table}
%

\bea
  \label{eqn:16}
F_1^- \es  - { e^{m_-^2/M^2} \over \lambda_- ( m_-  + m_+) }
\Big[ \Pi_2^B + ( m_1  + m_+ ) \Pi_3^B \Big]\,, \nnb \\
F_2^- \es
 { e^{m_-^2/M^2} \over \lambda_- ( m_-  + m_+ ) }
\Big[ \Pi_1^B + 2 m_1  \Pi_3^B - (m_1 - m_+ ) \Pi_4^B  \Big]\,, \nnb \\
G_1^- \es
 { e^{m_-^2/M^2} \over \lambda_- ( m_-  + m_+ ) }
\Big[ \Pi_6^B - ( m_1 - m_+ ) \Pi_7^B \Big]\,, \nnb \\
G_2^- \es
 - { e^{m_-^2/M^2} \over \lambda_- ( m_-  + m_+ ) }
\Big[ \Pi_5^B + 2 m_1  \Pi_7^B + ( m_1 + m_+)  \Pi_8^B \Big]\,, \nnb 
\eea
where $\Pi_i$ are the invariant functions in the Lorentz structures that are listed in Table \ref{tab:2} and $\Pi_i^B$ denotes the Borel transformed invariant function. 

The form factors $F_3^-,F_4^-,G_3^-$ and $G_4^-$ are equal to zero since the
corresponding Lorentz structures are absent in the theoretical part. Hence,
only the form factors $F_1^-,F_2^-,G_1^-$ and $G_2^-$ are needed in further
analysis. Note also that this result agrees with the HQET prediction in low recoil region~\cite{Chua:2019yqh}. To study the heavy hadron decay form factors at large recoil limit the soft-collinear effective can be applied~\cite{Mannel:2011xg,Wang:2015ndk}.

The Borel transformation and continuum subtraction procedure
is performed in the theoretical part with the help of the formula,
\bea
  \label{eqn:17}
\int_0^\infty d\sigma \frac{\rho (\sigma)}{{[{(p^{\prime 2} - s(\sigma)}]^n}}
&\to& \int_0^{\sigma_0} d\sigma \Bigg\{ {(-1)}^n \frac{e^{-s(\sigma)}
I(\sigma)}{{(n-1)}!{(M^2)}^{n-1}}\Bigg\} \nnb \\
\ek \Bigg[\frac{{(-1)}^{n-1}}{{(n-1)!}} e^{-s(\sigma)/M^2} \sum_{j=1}^{n-1}
\frac{1}{{{(M)}^2}^{n-j-1}} \frac{1}{s'}{(\frac{d}{d\sigma} \frac{1}{s'})}^{j-1}
I_n \Biggr|_{\sigma = \sigma_0}\Bigg]\,,
\eea
where
\bea
 I_n = \frac{\rho(\sigma)}{\bar{\sigma}}\,, \nnb 
\eea
and $\sigma_0$ is the solution of the equation $s = s_{th}$,
\bea
\sigma_0 = \frac{(s_{th} + m_1^2 - q^2) + \sqrt{{(s_{th} +
m_1^2 - q^2)}^2 - 4m_1^2(s_{th} - m_c^2)}}{2 m_1^2}\,, \nnb
\eea
in which $s_{th}$ is the continuum threshold.

At the end of this section, we calculate the semileptonic
decay widths $H_b \to H_c  l \nu  (l = e, \mu, \tau)$. To determine the decay width, we used the helicity amplitudes formalism.

In this formalism, the helicity amplitudes $\mathcal{H}_{\lambda_2,\lambda_w}$
are expressed in terms of vector and axial form factors, where
$\lambda_2 = \pm 1/2, \pm 3/2$ and $\lambda_w = \pm 1, 0$ are the
helicity components of the final baryon and vector meson.

The helicity amplitudes are determined as follows~\cite{Gutsche:2017wag}:.
\bea
  \label{eqn:18}
    \mathcal{H}_{\lambda_2, \lambda_w} = \epsilon^{+\mu}(\lambda_w) \lla
H_c (p', \lambda_2) \ve \Bar{c}
\gamma_\mu(1-\gamma_5)b \ve H_b(p_1,\lambda_1)\rra\,,
\eea

\bea
  \label{eqn:19}
\mathcal{H}_{1/2, t}^{V(A)}\es \pm \sqrt{\frac{2}{3} \frac{Q_\pm}{q^2}}
\frac{Q_\pm}{2 m_1 m_{-}} \Bigg\{ F_4^- (G_4^-) m_1 \nnb \pm F_1^- (G_1^-) M_\pm \nnb \\
\ar \Bigg[\frac{m_1}{m_{-}} F_3^- (G_3^-) + F_2^- (G_2^-)\Bigg] \frac{m_1^2 - m_{-}^2 - q^2}{2m_1} +
F_2^- (G_2^-) \frac{q^2}{m_1} \Bigg\}\,,\nnb \\
            \mathcal{H}_{1/2, 0}^{V(A)}\es \pm \sqrt{\frac{2}{3} \frac{Q_\pm}{q^2}}
\Bigg\{ F_4^- (G_4^-) \frac{m_1^2 - m_{-}^2 - q^2}{2m_{-}} \pm F_1^- (G_1^- )
\frac{Q_\pm  (M_{\pm})}{2m_1 m_{-}} \nnb \\ 
\ar \Bigg[\frac{m_1}{m_{-}} F_3^- (G_3^-) + F_2^- (G_2^-) \Bigg]
\frac{\vel \vec{p^\prime} \ver^2}{2m_{-}}  \Bigg\}\,, \nnb \\
            \mathcal{H}_{1/2, 1}^{V(A)}\es \sqrt{\frac{Q_\pm}{3}} \Bigg[ F_4^- (G_4^-) -
F_1^- (G_1^-) \frac{Q_\pm}{m_1 m_{-}} \Bigg]\,, \nnb \\
            \mathcal{H}_{3/2, 1}^{V(A)}\es \sqrt{Q_\pm} \Big[ F_4^- (G_4^-) \Big]\,, \nnb  \\
            \mathcal{H}_{-\lambda_2, -\lambda_w}^V\es \mathcal{H}_{\lambda_2, \lambda_w}^V\,, \nnb \\
            \mathcal{H}_{-\lambda_2, -\lambda_w}^A\es -\mathcal{H}_{\lambda_2, \lambda_w}^A\,,
\eea
where
\bea
M_\pm \es m_1 \pm m_{-}\,, \nnb \\
Q_\pm \es M_{\pm}^2 - q^2 \,. \nnb \\
|\vec{p^\prime}| \es \frac{\lambda^{1/2} (m_1^2, m_{-}^2, q^2)}{2m_1}\,. \nnb \\
\lambda(a,b,c) \es a^2 + b^2 + c^2 - 2 a b - 2 a c - 2 b c 
\eea
Note that, the above formula for the helicity amplitudes is derived for the general $\frac{1}{2}^+ \to \frac{3}{2}^-$ transitions. In our case, $F_3^-$,$F_4^-$,$G_3^-$, and $G_4^-$ are equal to zero. 

Using the helicity amplitudes, we get the differential decay widths for the corresponding transitions
\bea
  \label{eqn:20}
\frac{d\Gamma}{d{q^2}} = \frac{G^2 \vel V_{cb} \ver^2 \sqrt{Q_+ Q_-}
q^2 (1-\hat{m_\ell}^2)}{384 \pi^3 m_1^3} \Bigg[ (1+\frac{\hat{m_\ell}^2}{2}) \mathcal{H}_1 +
\frac{3}{2} \hat{m_\ell}^2 \mathcal{H}_2 \Bigg]\,,
\eea
where
\bea
\mathcal{H}_1\es \sum_{\lambda_2 = \pm 1/2 , \pm 3/2} \sum_{\lambda_w = \pm 1, 0}
\vel \mathcal{H}_{\lambda_2\lambda_w} \ver^2\,, \nnb \\
\mathcal{H}_2\es \sum_{\lambda_2 = \pm 1/2} \vel \mathcal{H}_{\lambda_2, t} \ver^2\,, \nnb \\
\mathcal{H}_{\lambda_2\lambda_w}\es \mathcal{H}_{\lambda_2\lambda_w}^V -
\mathcal{H}_{\lambda_2\lambda_w}^A\,, \nnb\\
\hat{m_\ell}^2\es \frac{m_\ell^2}{q^2}\,.\nnb
\eea

In the following section, we perform a numerical analysis of the obtained sum rules for the form factors as well as corresponding decay widths.

\section{Numerical analysis}
\label{sec:3}
This section is devoted to the numerical analysis of the sum rules obtained in the previous section. To perform the numerical analysis, the values of input parameters, which are collected in Table~\ref{tab:3}, are needed. 
\begin{table}[hbt]
\begin{adjustbox}{center}
\renewcommand{\arraystretch}{1.2}
\setlength{\tabcolsep}{6pt}
  \begin{tabular}{ll}
    \toprule
$m_c (\mu = m_c) $ = $ 1.27 \pm 0.02~\rm{GeV}$ \cite{PDG:2022pth}                       & \\
$m_{\Lambda_b} = 5619 \pm 0.17~\rm{MeV}$~\cite{PDG:2022pth}                     &
$m_{\Xi_b} = 5797 \pm 0.6~\rm{MeV}$~\cite{PDG:2022pth}                            \\
$m_{\Lambda_c^-} = 2625 \pm 0.19~\rm{MeV}$~\cite{PDG:2022pth}                   &
$m_{\Xi_c^-} = 2815 \pm 0.25~\rm{MeV}$~\cite{PDG:2022pth}                         \\
$m_{\Lambda_c^+} = 2860 \pm 0.28 ~\rm{MeV}$~\cite{PDG:2022pth}                  &
$m_{\Xi_c^+} = 2645 \pm 0.2~\rm{MeV}$~\cite{PDG:2022pth}                          \\
$f_1^{\Lambda_c} = 0.022 \pm 0.001~\rm{GeV}^3$~\cite{Groote:1996em}             &
$f_1^{\Xi_c} = 0.026 \pm 0.001~\rm{GeV}^3$~\cite{Groote:1996em}                                        \\
$f_2^{\Lambda_c} = 0.022 \pm 0.001~\rm{GeV}^3$ ~\cite{Groote:1996em}            &
$f_2^{\Xi_c} = 0.026 \pm 0.001~\rm{GeV}^3$~\cite{Groote:1996em}                                        \\
$\lambda_{\Lambda_c^-} = 0.050 \pm 0.005~\rm{GeV}^5$ (This work)                &
$\lambda_{\Xi_c^-} = 0.060 \pm 0.005~\rm{GeV}^5$ (This work)                      \\
$\tau_{\Lambda_b} = (1.471 \pm 0.009) \times 10^{-12}\,s$~\cite{PDG:2022pth}  &
$\tau_{\Xi_b} = (1.57 \pm 0.04) \times 10^{-12}\,s$~\cite{PDG:2022pth}          \\
$V_{cb} = (40.8 \pm 1.4) \times {10}^{-3}$~\cite{PDG:2022pth}              &
$V_{ud} = 0.973 \pm 0.004$~\cite{PDG:2022pth}                                \\
    \bottomrule
  \end{tabular}
\end{adjustbox}
  \caption{Numerical values of the input parameters appearing in the
  sum rules of the form factors, and branching ratio calculations. For the mass of charm quark, we used the value in $\overline{MS}$ scheme.}
  \label{tab:3}
\end{table}


It should be noted that in the numerical analysis, we neglect ${\cal O}(\alpha_s)$ corrections. Hence, to be consistent, we used the values of $f^{(1)}$ and $f^{(2)}$ for $\Lambda_b$ baryon obtained without ${\cal O}(\alpha_s)$ corrections \cite{Groote:1996em}.
When the SU(3) symmetry violation is taken into account, we obtain $f^{(1)} = f^{(2)} = 0.026 \pm 0.001$ for the $\Xi_b$ baryon.

In this study, we also calculated the residues
$\lambda_{\Lambda_c^-}$ and $\lambda_{\Xi_c^-}$ of the negative parity 
spin ${3\over 2}^-$ baryons, respectively, whose
values are presented in Table \ref{tab:3}.   

In addition to these input parameters, the sum rules involve
two more additional parameters; the Borel mass, $M^2$, and the continuum
threshold $s_{th}$. The working region of $s_{th}$ is determined by imposing the condition that the two-point sum rules  predict the mass of the baryon within an accuracy range of $5-10\%$ compared to the experimental value. With this condition, we obtain the following working region for $s_{th}$, $10~\rm{GeV^2} \leq s_{th} \leq 12~\rm{GeV^2}$.

The working region of $M^2$ is determined with the help of two requirements:
\begin{itemize}[nosep]
    \item[a)] Higher-order twist contributions should be suppressed compared to the leading twist one.
    \item[b)] The contributions from the continuum and higher states should constitute $40\%$ of the total result. Our numerical analysis shows that both conditions are satisfied in the region  $2.5~GeV^2 \leq M^2 \leq 4~GeV^2$.
\end{itemize}

After presenting the values of all input parameters and determining the working regions of $M^2$ and $s_{th}$, we can proceed 
with calculations of the form factors associated with the $\Xi_b \to \Lambda_c (2625)$ and $\Xi_b \to \Xi_c (2815)$ transitions.

To calculate the semileptonic decay widths $\Xi_b \to \Lambda_c (2625) l \nu$
and $\Xi_b \to \Xi_c (2815) l \nu$, we need to know the $q^2$ dependency of all the form factors. It is important to note that the LCSR predictions are reliable only in the region $q^2 \lesssim 6~GeV^2$. To extend the results to the whole physical region
$m_e^2 \leq q^2 \leq {(m_{1}-m_{-})}^2$, we will apply z-series parametrization for the form factors (see \cite{Bourrely:2008za}).
\bea
\label{eqn:21}
F_i(q^2)  = {1\over 1- \ds{q^2 \over m_{pole}^2}}
\Big\{f_i(0) + \alpha [ z(q^2) - z(0) ] + 
\beta [ z(q^2) - z(0) ]^2 \Big\} \,.
\eea
where
\bea
    z(t) = \frac{\sqrt{t_+ - q^2} - \sqrt{t_+ - t_0}}{\sqrt{t_+ - q^2} +
\sqrt{t_+ - t_0}}\,,\nnb
\eea
with 
\bea
t_\pm \es (m_1 \pm m_{-})^2\,, \nnb \\
 t_0  \es t_+ \Bigg(1 - \sqrt{1- \frac{t_-}{t_+}}\Bigg)\,.\nnb
 \eea
The mass of the resonances for $b \to c$ transitions are:
\begin{equation*}
    \begin{aligned}
    m_{pole} =
        \begin{cases}
            6.275~\rm{GeV} &f_1 ,\\
            6.33~\rm{GeV} &f_2,\\
            6.706~\rm{GeV} &g_1,\\
            6.741~\rm{GeV} &g_2.
        \end{cases}
    \end{aligned}
  \end{equation*}
The fitting parameters, $f_i(0)$, $\alpha$ and $\beta$ are presented in Table~\ref{tab:4}.  
\begin{table}[hbt]
  \renewcommand{\arraystretch}{1.5}
\setlength{\tabcolsep}{6pt}
\begin{adjustbox}{center}
\begin{tabular}{cccc|ccc}
  \toprule
& \multicolumn{3}{c|}{$\Lambda_b \to \Lambda_c$} & 
  \multicolumn{3}{c}{$\Xi_b \to \Xi_c$} \\
        & $f_1(0)$ &  $\alpha$  & $\beta$ 
        & $f_1(0)$ &  $\alpha$  & $\beta$ \\
  \midrule
$F_1^-$ & $-1.00 \pm 0.15$  & $4.29$  & $16.80$ & $-1.04 \pm 0.14$  & $ 5.67$ & $ 7.25$  \\
$F_2^-$ & $ 0.37  \pm 0.06$ & $-4.38$ & $19.30$ & $ 0.38 \pm 0.06$  & $-5.04$ & $26.37$  \\
$G_1^-$ & $-0.63 \pm 0.10$  & $ 0.50$ & $32.30$ & $-0.66 \pm 0.11$  & $ 1.25$ & $29.24$  \\
$G_2^-$ & $ 0.37 \pm 0.06$  & $-4.70$ & $23.41$ & $ 0.38 \pm 0.06$  & $-5.38$ & $30.87$  \\
\bottomrule
\end{tabular}
\end{adjustbox}
  \caption{Fit parameters for the form factors of the $\Lambda_b \to
\Lambda_c(2815) \ell \nu_\ell$ transition at $M^2=3\,GeV^2,~s_0 = 10\,GeV^2$,
and $\Xi_b \to \Xi_c(2625) \ell \nu_\ell$ transition at $M^2=3\,GeV^2,~s_0 = 11\,GeV^2$.}
  \label{tab:4}
\end{table}

The errors in the form factors due to the variation of $M^2$ and $s_{th}$ in their working regions, as well as the uncertainties in the input parameters, are taken into account for the obtained results. All the uncertainties are taken into account quadratically.

Using the lifetime of the $H_b$ and the results of the form factors, one can easily calculate the corresponding branching ratios of the semileptonic $H_b \to H_c \ell \nu$ decays presented in Table~\ref{tab:5}. 

Finally, the obtained results for the form factors enable us to evaluate the decay widths
of the color-allowed two body nonleptonic decays $H_b \to H_c M$, 
where $M$ corresponds to  pseudoscalar $\pi^-$, or  vector meson $\rho^-$. To calculate the decay widths of these nonleptonic decays, we need the values of the form factors at the point $q^2 = m_M^2$, where
$m_M$ is the mass of vector or pseudoscalar mesons.

The straightforward calculation leads to the following result for the
nonleptonic decay width
\bea
 \label{eqn:22}
    \Gamma(H_b \to H_c M) = \frac{G_F^2 f_M^2 \abs{\vec{p^\prime}}}{32 \pi m_1^2}
\abs{V_{cb} V_{uq}}^2 a_1^2 \mathcal{H}_2^2 m_M^2\,,
\eea
where $f_{M}$ is the leptonic decay constant of the corresponding meson, $N_c$ is the color factor, $c_1=-0.25$, $c_2 = 1.1$ \cite{Chua:2018lfa}, and $a_1 = c_1 + \ds{ c_2\over N_c}$ \cite{Hsiao:2019ann}. The results for the branching ratios of the nonleptonic $H_b \to H_c M$ decay widths are also presented in \ref{tab:5}. For comparison, we also demonstrate the results from other approaches. We would like to note that the decay width is obtained within the naive factorization approximation and the non-factorizable contributions as well as the scale dependency are neglected. The calculations of the non-factorizable contributions have been discussed widely in the framework of different approaches. (See for example~\cite{Ali:2007ff,Ali:1998eb,Rui:2022sdc,Cheng:1996cs,Ivanov:1997hi} and references therein.)

The experimental result on branching ratios is available only for the $\Lambda_b \to \Lambda_c(2625) \ell \nu_\ell$ decay. Our prediction of the branching ratio for this decay is quite compatible with this data.

From Table~\ref{tab:5}, we deduce that our results for the branching ratios of the semileptonic decay modes
$\Lambda_b \to \Lambda_c(2625) \ell \nu_\ell$ and $\Xi_b \to \Xi_c(2815) \ell \nu_\ell$  
are in good agreement with the results of the light-front approach \cite{Li:2022hcn}. However, they are considerably
different from the results predicted within the confined covariant quark model (CCQM) \cite{Gutsche:2018nks}, heavy quark spin symmetry (HQSS) \cite{Nieves:2019kdh}, and constituent quark model (CQM) \cite{Pervin:2005ve}.

On the other hand, our predictions on the branching ratios of the nonleptonic decays $\Lambda_b(\Xi_b)  \to \Lambda_c(\Xi_c) \pi$ are in good agreement with the findings of LFQM~\cite{Li:2022hcn,Chua:2018lfa}. However, our predictions on the widths of
the $\Lambda_b(\Xi_b)  \to \Lambda_c(\Xi_c) \rho$ are approximately 2.5 times larger than those predicted by the LFQM.

 Naively, one expects that the ratio,
\bea
R = {Br[ \Lambda_b(\Xi_b)  \to \Lambda_c(\Xi_c) \rho] \over 
Br[\Lambda_b(\Xi_b)  \to \Lambda_c(\Xi_c) \pi]}\,,\nnb
\eea
should approximately be equal to 3 due to the three polarization states of the $\rho$ meson. In our case, this ratio is equal to $\sim 2.5$. The difference can be attributed to the mass difference of the $\rho$ and $\pi$ mesons.

Our analysis shows that the branching ratios of the semileptonic decays are slightly larger
than 1\%, which could be  accessible in experiments planned to be
conducted at LHCb in the near future. Measurement of the studied decays can
provide useful information about the inner structures of $\Lambda_c (2625)$
and $\Xi_c (2815)$ baryons. 


\begin{table}[hbt]
\begin{adjustbox}{width=\columnwidth,center}
  \renewcommand{\arraystretch}{1.6}
\setlength{\tabcolsep}{6pt}
\begin{tabular}{lccccccc}
  \toprule
  \multicolumn{1}{c}{ Decay }                                         &
  \multicolumn{1}{c}{ This Study} & \multicolumn{1}{c}{ Experiment~\cite{PDG:2022pth} } &
  \multicolumn{1}{c}{ LFQM~\cite{Li:2022hcn} }     &
  \multicolumn{1}{c}{ CCQM~\cite{Gutsche:2018nks} }&
  \multicolumn{1}{c}{ HQSS~\cite{Nieves:2019kdh} } &
  \multicolumn{1}{c}{ CQM~\cite{Pervin:2005ve} }   &
  \multicolumn{1}{c}{ LFQM~\cite{Chua:2018lfa} }   \\
  \midrule
$\Lambda_b^0 \rightarrow \Lambda_c^{+} e^{-} \nu_e$    & $1.44 \pm 0.56$             &
$-$                      & $1.653 \pm 0.114$                      & $0.17 \pm 0.03$  &
$-$                      & $(0.88-1.40)$               & $-$                         \\
$\Lambda_b^0 \rightarrow \Lambda_c^{+} \mu^{-} v_\mu$  & $1.42  \pm 0.56 $           &
$1.3_{-0.5}^{+0.6}$      & $1.641 \pm 0.113$                      & $0.17 \pm 0.03$  &
$3.5_{-1.2}^{+1.3}$      & $(0.88-1.40)$               & $-$                         \\
  $\Lambda_b^0 \rightarrow \Lambda_c^{+} \tau^{-} v_\tau$ & $0.11 \pm 0.04 $         & $-$  &
$0.1688 \pm 0.0116$                    & $0.018 \pm 0.004$           & $0.38_{-0.08}^{+0.09}$ &
$(0.18-0.22)$            \\
  \midrule
  \midrule
$\Xi_b^{0(-)} \rightarrow \Xi_c^{+(0)} e^{-} v_e$       & $1.55 \pm 0.62 $      & $-$  &
$1.698 \pm 0.122(1.803 \pm 0.132)$     & $-$            & $-$                   & $-$  & $-$ \\
$\Xi_b^{0(-)} \rightarrow \Xi_c^{+(0)} \mu^{-} v_\mu$   & $1.50 \pm 0.60$       & $-$  &
$1.685 \pm 0.121(1.789 \pm 0.131)$     & $-$            & $-$                   & $-$  & $-$ \\
$\Xi_b^{(-)} \rightarrow \Xi_c^{+(0)} \tau^{-} v_\tau$  & $0.12 \pm 0.048$      & $-$  &
$0.1758 \pm 0.0126(0.1868 \pm 0.0137)$ & $-$            & $-$                   & $-$  & $-$ \\
  \midrule
  \midrule
$\Lambda_b^{0} \rightarrow \Lambda_c^{+} \pi^-$         & $0.40 \pm 0.16$       & $-$  &
$0.310 \pm 0.015$     & $-$                             & $-$        & $-$   &
$\left(2.40_{-1.82}^{+4.09}\right)\times 10^{-1}$           \\
$\Lambda_b^{0} \rightarrow \Lambda_c^{+} \rho^-$        & $1.1 \pm 0.4$         & $-$  &
$0.450 \pm 0.023$     & $-$                             & $-$        & $-$   &
$\left(4.38_{-3.17}^{+6.78}\right)\times 10^{-1}$           \\
  \midrule
  \midrule
$\Xi_b^{0(-)} \rightarrow \Xi_c^{+(0)} \pi^-$         & $0.43 \pm 0.16 $        & $-$  &
$0.310 \pm 0.017$     & $-$                         & $-$        & $-$  &
$\left(3.32_{-2.85}^{+6.08}\right)\times 10^{-1}$               \\
$\Xi_b^{0(-)} \rightarrow \Xi_c^{+(0)} \rho^-$     & $1.1 \pm 0.4$              & $-$  &
$0.430 \pm 0.025$     & $-$                         & $-$        & $-$   &
$\left(6.10_{-4.84}^{+9.95}\right)\times 10^{-1}$              \\
\bottomrule
\end{tabular}
\end{adjustbox}
  \caption{The branching ratios of the
$\Lambda_b \to \Lambda_c(2625) \ell \nu_\ell$,
$\Xi_b \to \Xi_c(2815) \ell \nu_\ell$,
$\Lambda_b \to \Lambda_c(2625) \pi(\rho)$, and
$\Xi_b \to \Xi_c(2815) \pi(\rho)$
decays. Here all numerical values are presented in percent (\%).}
  \label{tab:5}
\end{table}



\section{Summary}
\label{sec:conclusion}

In the present work, we first calculate the form factors for
${1/2}^+ \to {3/2}^-$, i.e., $\Lambda_b \to \Lambda_c (2625)$ and $\Xi_b \to \Xi_c (2815)$
transitions within the light-cone sum rules method by using the DAs of the
$\Lambda_b$ and $\Xi_b$ baryons. Having  obtained the form factors, we estimate the branching ratios of
semileptonic $\Xi_b \to \Xi_c(2815) \ell \nu_\ell$, $\Lambda_b \to \Lambda_c(2625) \ell \nu_\ell$
as well as nonleptonic $\Xi_b \to \Xi_c(2815) \rho(\pi)$ and $\Lambda_b \to \Lambda_c(2625) \rho(\pi)$

Within the sum rules accuracy, our result on the branching ratio for the semileptonic $\Lambda_b \to \Lambda_c (2625)$ decay is in good agreement with the existing experimental data.  We also compared our findings with the predictions of other approaches.

Relatively large branching ratios, which follow from our calculations, indicate that, hopefully, considered semileptonic decays would be measured in future experiments to be carried out at LHCb.



 \bibliographystyle{utcaps_mod}
 \bibliography{all}



\end{document}